\documentclass[sigconf, nonacm]{acmart}

\AtBeginDocument{%
  }

\setcopyright{acmlicensed}
\copyrightyear{2026}
\acmYear{2026}
\acmDOI{XXXXXXX.XXXXXXX}

\acmISBN{978-1-4503-XXXX-X/2018/06}

\usepackage{multirow}
\usepackage{graphicx}
\usepackage[table,xcdraw]{xcolor}

\usepackage{pifont}
\newcommand{\xmark}{\ding{55}}%

\usepackage{listings}
\usepackage{tcolorbox}

\colorlet{punct}{red!60!black}
\definecolor{background}{HTML}{EEEEEE}
\definecolor{delim}{RGB}{20,105,176}
\definecolor{codegray}{rgb}{0.5,0.5,0.5}
\colorlet{numb}{magenta!60!black}

\lstdefinelanguage{json}{
    basicstyle=\normalfont\ttfamily,
    numberstyle=\scriptsize,
    stepnumber=1,
    numbersep=8pt,
    showstringspaces=false,
    breaklines=false,
    frame=single,
    backgroundcolor=\color{background},
    moredelim=[is][\color{black}]{<<}{>>},
    moredelim=[is][\color{codegray}]{<-}{->},
}

\lstdefinestyle{promptstyle}{
  basicstyle=\ttfamily\footnotesize,
  backgroundcolor=\color{white},
  frame=single,
  rulecolor=\color{black},
  framerule=0.5pt,
  framesep=5pt,
  linewidth=\columnwidth,
  xleftmargin=1pt,
  xrightmargin=1pt,
  breaklines=true,
  breakatwhitespace=true,
  breakindent=0pt,
  breakautoindent=false,
  columns=fullflexible,
  keepspaces=true,
  showstringspaces=false,
  numbers=none,
  aboveskip=0.8em,
  belowskip=0.8em
}

\newcommand{\name}{AutoMUD}

\usepackage{subcaption}

\usepackage[acronym,shortcuts]{glossaries-extra}
\glssetcategoryattribute{acronym}{nohyper}{true}
\setabbreviationstyle[acronym]{long-short}

\newacronym{ace}{ACE}{Access Control Entry}
\newacronym{acl}{ACL}{Access Control List}
\newacronym{cms}{CMS}{Cryptographic Message Syntax}
\newacronym{ietf}{IETF}{Internet Engineering Task Force}
\newacronym{iot}{IoT}{Internet of Things}
\newacronym{jsonl}{JSONL}{line-oriented JSON}
\newacronym{llm}{LLM}{Large Language Model}
\newacronym{mud}{MUD}{Manufacturer Usage Description}
\newacronym{pcap}{PCAP}{Packet Capture}
\newacronym{rag}{RAG}{Retrieval Augmented Generation}
\newacronym{yang}{YANG}{Yet Another Next Generation}

\usepackage{todonotes}

\usepackage{booktabs}
\usepackage{tabularx}
\usepackage{array}

\begin{document}

\title{From Source Code to Network Profile:\\ Automated and Traceable MUD Profile Generation for IoT Devices}

\author{Alessandro Lotto}
\orcid{}
\affiliation{%
  \institution{University of Padua}
  \city{Padua}
  \country{IT}
}
\email{alessandro.lotto@math.unipd.it}

\author{Abdulla R. A. Almenhali}
\orcid{}
\affiliation{%
  \institution{University of Padua}
  \city{Padua}
  \country{IT}
}
\email{abdulla.almenhali@studenti.unipd.it}

\author{Savio Sciancalepore}
\orcid{}
\affiliation{%
  \institution{Eindhoven University of Technology}
  \city{Eindhoven}
  \country{NL}
}
\email{s.sciancalepore@tue.nl}

\author{Alessandro Brighente}
\orcid{}
\affiliation{%
  \institution{University of Padua}
  \city{Padua}
  \country{IT}}
\email{alessandro.brighente@unipd.it}

\author{Mauro Conti}
\orcid{}
\affiliation{%
 \institution{University of Padua}
 \city{Padua}
 \country{IT}
}
\email{mauro.conti@unipd.it}

\renewcommand{\shortauthors}{Lotto et al.}

\begin{abstract}

The Manufacturer Usage Description (MUD) standard allows manufacturers of Internet of Things (IoT) devices to define in a MUD file the profile of the expected network behavior of their devices. The MUD file can then be translated into enforceable access-control policies, thus restricting compromised devices to operate solely through manufacturer-defined communication patterns.
However, the practical adoption of MUD depends on profiles that are accurate, complete, and maintainable.
Existing approaches use traffic-based automation to reduce the manual effort to create MUD profiles, but require device deployment and prolonged monitoring, while capturing only behavior exercised during observation.
Rare, failure-triggered, or configuration-dependent communications may remain absent, producing incomplete policies that may disrupt legitimate operation and offer limited insight into the software components responsible for each rule.

In this paper, we present \name, a source-code-driven tool that generates traceable MUD profiles for IoT devices from their firmware and software source code.
\name\ combines static and syntactic extraction, retrieval-grounded language-model reasoning, and deterministic validation and compilation to recover the communication behavior characterizing an IoT device and translate eligible endpoints into policy rules.
By analyzing code-level evidence, \name\ exposes even rarely exercised and conditional communication paths, links every generated rule to its source-level provenance, and preserves excluded findings with explicit reasons for review.
Our evaluation on a Linux-based repository demonstrates that \name\ recovers the complete communication behavior, consolidates the validated behavior into semantic endpoint groups, and generates the expected structurally valid MUD profile.
Through a controlled semantic fault-injection campaign, we also demonstrate that 
\name\ enables an analyst to detect, localize, explain, and correct propagated errors, recovering policies semantically identical to their clean counterparts. 
\end{abstract}




\keywords{}


\maketitle

\section{Introduction}  \label{sec:Introduction}
\ac{iot} devices increasingly support automation, sensing, and remote control across homes, healthcare, industry, energy systems, and urban infrastructure~\cite{Alrawi:2019:SoK, Yu:2020:IoTSurvey}.
At the same time, their widespread deployment, combined with their constrained resources, limited monitoring, and weak update practices, expands the attack surface of modern networks.
Once compromised, an \ac{iot} device may communicate with malicious endpoints, ports, or services unrelated to its intended function~\cite{Dritsas:2025:IoTCyberSurvey}.
Restricting each device to the communications required for its legitimate operation can therefore reduce the impact of compromise.

The \ac{mud} standard of the \ac{ietf} implements this principle by allowing manufacturers to describe the intended network behavior of a device through a machine-readable \ac{mud} profile~\cite{MUD-rfc}.
A \ac{mud} controller can retrieve this profile and translate it into access-control policies, treating undeclared traffic as unauthorized\cite{MUD-rfc, HernandezRamos:2021:MUDSurvey, Mazhar:2021:MUDSurvey}. 
Despite its potential, \ac{mud} adoption today remains limited, especially because producing and maintaining accurate profiles requires detailed knowledge of devices' communication behavior~\cite{HernandezRamos:2021:MUDSurvey, Watrobski:2022:NISTIR8349, Mazhar:2021:MUDSurvey}.
Within large manufacturers of \ac{iot} devices, this knowledge may be distributed across firmware, cloud-service, integration, and network teams, making manual profile construction time consuming, error-prone, and difficult to maintain across devices, configurations, and software versions~\cite{HernandezRamos:2021:MUDSurvey, Watrobski:2022:NISTIR8349}.

To date, the only automated solution to this problem is the use of traffic-based tools, such as MudGee~\cite{hamza2022_tdsc}, which reduce manual effort of \ac{mud} profile generating by translating observed communications into \ac{mud} rules.
However, they require devices to be deployed and monitored for sufficiently long periods to obtain representative traces~\cite{Hamza:2018:ClearAsMUD, Matheu:2019:MUDProfiles, Hamza:2018:MUDgeeTool}.
Even then, rare, configuration-dependent, update-related, or failure-triggered communications may remain unobserved and consequently absent from the generated policy~\cite{Alrawi:2019:SoK, Yu:2020:IoTSurvey, Dritsas:2025:IoTCyberSurvey}.
Such omissions may later disrupt legitimate operation when the profile is enforced.
Moreover, if the device is compromised during data collection, its \ac{mud} profile could inevitably encode malicious behavior.
Furthermore, traffic traces may show that some communications occurred, but they provide limited evidence about the software component, configuration, or execution path that caused it.
In this context, manufacturers could significantly benefit from an automated \ac{mud} generation approach that can recover the expected network behavior by connecting each communication pattern to inspectable evidence without requiring device deployment.

{\bf Contribution.} In this paper, we present \name, a source-code-driven tool for automated and traceable \ac{mud} profile generation.
\name\ builds on the observation that firmware and software artifacts contain direct evidence of implemented communication behavior, including networking APIs, protocol handlers, configuration structures, and connection-setup logic.
Building on such rationale, \name\ processes a source code-base through lightweight static and syntactic code extraction, semantic retrieval, retrieval-grounded \ac{llm} reasoning, and deterministic validation and compilation. 
\name\ isolates network-relevant applications, discovers protocol and endpoint candidates, validates their values and uses, infers communication roles and transport semantics, and consolidates eligible behaviors into \ac{mud} rules.
Each generated rule remains linked to the source evidence and intermediate decisions that produced it, while unresolved or non-representable findings are retained for review and not converted into permissions.
Since \name\ analyzes implemented capabilities, it can generate a source-grounded \ac{mud} profile before device deployment and without requiring relevant communications to appear in traffic traces.
This enables \name\ to recover even rare, configuration-dependent, and difficult-to-exercise communication paths that may remain unobserved even during prolonged monitoring.
In addition, its source-to-rule traceability provides manufacturers with inspectable evidence for validating and maintaining the resulting policy.
We evaluate \name\ on a Linux-based repository containing curl, wget, and OpenSSH services, which represent heterogeneous networking implementations, demonstrating the effectiveness of the tool and its potential for 
detecting and correcting propagated semantic errors.
In summary, this paper makes the following contributions:
\begin{itemize}
    \item \textbf{Automated source-code-driven \ac{mud} generation.}
    We design \name\ to recover implemented network behavior from static firmware and software artifacts and generate a \ac{mud} profile without requiring device deployment.
    This approach further exposes rare and configuration-dependent behavior that may be missed during time-limited observation campaigns.
    
    \item \textbf{Traceable protocol-to-policy synthesis.}
    We introduce a systematic procedure that maps source evidence into \ac{mud}-level protocol, endpoint, transport, role, and direction abstractions.
    Intermediate artifacts preserve the provenance of generated rules and the reasons associated with excluded candidates, supporting review of both policy inclusions and omissions.
    
    \item \textbf{Source-grounded evaluation and semantic-accountability assessment.}
    We evaluate \name\ on several controlled Linux-based codebases, and show that it reconstructs the expected profile while 
    preventing non-endpoint values from becoming permissions.
    Through controlled semantic fault injection tests, we further demonstrate that its traceability artifacts can expose, localize, explain, and support the correction of an erroneous rule that structural validation alone does not detect.
\end{itemize}

{\bf Roadmap.} The remainder of the paper is structured as follows.
Sec.~\ref {sec:Background} introduces preliminary notions. 
Sec.~\ref{sec:RelatedWork} reviews the current literature. 
Sec.~\ref{sec:Methodology} presents the \name\ tool. 
Sec.~\ref{sec:ExpEval} presents the experimental evaluation and results. 
Sec.~\ref{sec:Discussion} discusses our key findings. 
Finally, Sec.~\ref{sec:Conclusions} concludes the paper.

\section{The MUD Standard}   \label{sec:Background}
\textbf{MUD Architecture.}
\ac{mud} is an \ac{ietf} standard designed to reduce the attack surface of \ac{iot} devices~\cite{MUD-rfc}.
It builds on the observation that many such devices have a narrow operational purpose and therefore require only a limited set of network interactions~\cite{Mazhar:2021:MUDSurvey}.
According to \ac{mud}, manufacturers may encode these interactions as explicit access-control policies in a cryptographically signed \ac{mud} file, which is published on a \ac{mud} server~\cite{HernandezRamos:2021:MUDSurvey}.
When joining a network, an \ac{iot} device announces a \ac{mud} URL identifying where to find its \ac{mud} profile.
A \ac{mud} controller deployed by the user is responsible for retrieving the file, verifying its signature, and translating its contents into enforceable network policies.
Communication patterns not included in the profile are treated as unauthorized~\cite{MUD-rfc,HernandezRamos:2021:MUDSurvey,Mazhar:2021:MUDSurvey}.

\textbf{MUD File.}
A \ac{mud} file is a JSON serialization of YANG modules describing the communication requirements of an \ac{iot} device~\cite{MUD-rfc}.
Its \ac{mud} section holds profile metadata and references to the applicable policies, while its \ac{acl} container defines the permitted communications.
Each \ac{acl} comprises a set of \acp{ace}, each combining rule metadata with a matching block and an action.
The matching block specifies the conditions under which the rule applies, including the IP family, network-layer protocol, transport-layer ports, and endpoint abstractions.
The associated action determines whether matching traffic is accepted or dropped.
Endpoint abstractions represent categories of peers, such as devices from the same manufacturer or manufacturer-defined services, avoiding dependence on fixed network addresses.
The direction of communication is encoded by the policy container that references each \ac{acl}.
Rules under \texttt{from-device-policy} describe traffic originating from the \ac{iot} device, as opposite to rules under \texttt{to-device-policy}. 
\section{Related Work}  \label{sec:RelatedWork}
Generating and maintaining accurate \ac{mud} profiles remain significant obstacles to \ac{mud} wider adoption~\cite{HernandezRamos:2021:MUDSurvey,Watrobski:2022:NISTIR8349,Mazhar:2021:MUDSurvey}.
Manual construction requires network expertise and detailed knowledge of device behavior, making it costly to scale across devices, configurations, and software versions.
We summarize in Table~\ref{tab:Tool-comparison} the existing approaches for automated \ac{mud} file generation, and we briefly explain them below.

\begin{table*}[]
\centering
\caption{Comparison of approaches for automated \ac{mud} profile generation.}
\label{tab:Tool-comparison}
\renewcommand{\arraystretch}{1.15}
\resizebox{.85\textwidth}{!}{%
\begin{tabular}{l|c|c|c|c|c}
\hline
\rowcolor[HTML]{C0C0C0} 
\multicolumn{1}{c|}{\cellcolor[HTML]{C0C0C0}\textbf{Approach}} &
  \textbf{\begin{tabular}[c]{@{}c@{}}Evidence\\ Source\end{tabular}} &
  \textbf{\begin{tabular}[c]{@{}c@{}}Policy\\ operation\end{tabular}} &
  \textbf{\begin{tabular}[c]{@{}c@{}}Device\\ deployment \\ not required \end{tabular}} &
  \textbf{\begin{tabular}[c]{@{}c@{}}Behavior\\ visibility\end{tabular}} &
  \textbf{\begin{tabular}[c]{@{}c@{}}Rule\\ justification\end{tabular}} \\ \hline
\textit{MudGee~\cite{Hamza:2018:ClearAsMUD,Hamza:2018:MUDgeeTool}} &
  \begin{tabular}[c]{@{}c@{}}Passive traffic\\ traces\end{tabular} &
  \begin{tabular}[c]{@{}c@{}}Generate\\ new profile\end{tabular} &
  \xmark &
  \begin{tabular}[c]{@{}c@{}}Executed and observed\\ communications\end{tabular} &
  Traffic-flow evidence \\
\rowcolor[HTML]{EFEFEF} 
\textit{\ac{mud}-PD~\cite{NIST:2020:MUDPDTool}} &
  \begin{tabular}[c]{@{}c@{}}Traffic and\\ network context\end{tabular} &
  \begin{tabular}[c]{@{}c@{}}Generate\\ new profile\end{tabular} &
  \xmark &
  \begin{tabular}[c]{@{}c@{}}Observed communications in\\ their deployment context\end{tabular} &
  \begin{tabular}[c]{@{}c@{}}Flow and network-context\\ evidence\end{tabular} \\
\textit{Matheu et al.~\cite{Matheu:2019:MUDProfiles}} &
  \begin{tabular}[c]{@{}c@{}}Active security\\ testing\end{tabular} &
  \begin{tabular}[c]{@{}c@{}}Refine\\ existing profile\end{tabular} &
  \xmark &
  \begin{tabular}[c]{@{}c@{}}Behaviors exercised\\ during testing\end{tabular} &
  \begin{tabular}[c]{@{}c@{}}Test-observation\\ evidence\end{tabular} \\
\rowcolor[HTML]{EFEFEF} 
\textit{Ours - \name} &
  Source code &
  \begin{tabular}[c]{@{}c@{}}Generate\\ new profile\end{tabular} &
  \checkmark &
  \begin{tabular}[c]{@{}c@{}}Code-based implemented\\ communication paths\end{tabular} &
  \begin{tabular}[c]{@{}c@{}}Source-to-rule provenance\\ and exclusion evidence\end{tabular} \\ \hline
\end{tabular}%
}
\end{table*}

\textbf{Traffic-based generation.}
Existing literature has primarily derived \ac{mud} profiles from observed device communications.
\textit{MudGee}, proposed in~\cite{Hamza:2018:ClearAsMUD}, extracts communication patterns from packet traces and translates them into \ac{mud} rules. 
Similarly, \ac{mud}-PD~\cite{NIST:2020:MUDPDTool} combines captured traffic with contextual information, such as device identity and observed service interactions, to support profile construction and operational deployment.
Both approaches generate policies from behaviors exercised and observed during monitoring.
By contrast, \textit{Matheu et al.}~\cite{Matheu:2019:MUDProfiles} actively stimulate device functionality through security testing, using the resulting observations to refine an existing \ac{mud} profile rather than generate one from scratch.

Although being valuable tools for users of \ac{iot} devices, approaches based on device execution remain bounded by the behaviors exercised during observation. For example, MudGee requires traffic collected over four to six months to obtain representative traces~\cite{Hamza:2018:ClearAsMUD,Hamza:2018:MUDgeeTool}.
Nevertheless, rare, configuration-dependent, update-related, or failure-triggered communications may not occur during the collection period and may consequently be omitted from the generated policy.
Moreover, traffic identifies observed endpoints and flows but provides limited evidence about the software component, configuration, or execution path responsible for them.
This complicates determining whether a rule represents intended functionality, deployment-specific behavior, or test activity, particularly when profiles must be reviewed or maintained across software versions.

\textbf{Source-code analysis.}
Source code provides a complementary evidence base because it contains the implementation logic underlying device communications, including networking APIs, protocol handlers, constants, configuration structures, and conditional paths.
Static analysis can identify calls such as \texttt{socket()}, \texttt{connect()}, and \texttt{bind()} and track the propagation of ports, addresses, and configuration values~\cite{Nielson:1999:PPA,Stevens:2004:UNP,Kerrisk:2010:TLPI}.
Static and graph-based frameworks further demonstrate that precise program facts can be extracted from large repositories~\cite{Palix:2010:CoccinelleLinux,Yamaguchi:2014:CodePropertyGraphs}.

However, \ac{mud} generation requires more than just detecting API calls or numeric constants.
Low-level evidence must be interpreted as policy-level behavior, including the relevant protocol, transport, port, IP family, communication role, and direction.
Closing this semantic gap is particularly challenging in heterogeneous \ac{iot} repositories that combine different languages, libraries, build systems, and platform-dependent code~\cite{Dritsas:2025:IoTCyberSurvey,UlHaq:2023:FirmwareSurvey}, and no existing solutions are available for this scope.
Thus, additional reasoning is required to transform static code extraction into \ac{mud}-level abstractions.

\textbf{LLM-assisted interpretation.}
Recent work shows that \acp{llm} can support code understanding and program analysis by identifying relevant functions, reconstructing relationships across code fragments, and assisting data-flow reasoning~\cite{Li:2024:IRIS,Wang:2024:LLMDFA,Salih:2025:LLMStaticSurvey,Wang:2025:LLMProgAnalysis}.
These capabilities can complement static analysis when communication semantics are distributed across constants, configuration structures, API calls, and connection-handling logic.
Although previous surveys identify automated and human-assisted generation as important to reducing the cost of \ac{mud} adoption~\cite{HernandezRamos:2021:MUDSurvey,Mazhar:2021:MUDSurvey}, they do not provide a concrete source-to-policy generation procedure.
\name\ addresses this and the previously mentioned gaps by combining deterministic source extraction with retrieval-grounded \ac{llm} reasoning to recover \ac{mud}-relevant behavior, generate enforceable rules, and link both included and excluded findings to their supporting software evidence.

\section{AutoMUD}   \label{sec:Methodology}
This section presents \name, our source-code-driven tool for automated \ac{mud} profile generation.
Sec.~\ref{sec:reqs} defines its design assumptions and requirements, while Sect.~\ref{subsec:Overview}-\ref{subsec:explainability-validation} describe the processing pipeline and its output artifacts.
Sec.~\ref{sec:PromptEngineering} then summarizes the constraints imposed on \ac{llm} interactions.

\subsection{Design Considerations and Requirements} \label{sec:reqs}
We consider the use case of an \ac{iot} software manufacturer whose device software is available as a traversable source tree containing source files, headers, scripts, build files, and configuration artifacts. 
We make no assumptions regarding naming conventions on developer-defined networking functions.


Overall, our design is driven by four requirements.
    First, we should consider that \ac{iot} software may come in the form of heterogeneous, multi-language software trees 
    containing application code, libraries, scripts, configuration artifacts, and unrelated components. As repository size and complexity grow, monolithic indexing approaches may suffer from scalability and retrieval-efficiency limitations. This motivates application isolation and application-specific retrieval indexes, which reduce search space, improve retrieval relevance, and enable scalable repository analysis ({\bf R1 - Repository scalability}).
    Second, the tool must recover the abstractions required for policy generation.
    These include protocol or service identity, transport and IP family, port and endpoint semantics, device role and communication direction, and the conditions under which the behavior is reachable or active ({\bf R2 - MUD-level behavior recovery}).
    Moreover, the output of the tool must be a structurally valid \ac{mud} file, and only source-supported endpoint records with concrete, internally consistent packet-level semantics may become enforceable rules.
    Unresolved or non-representable findings must remain outside the automatically generated policy ({\bf R3 - Conservative policy synthesis}).
    Finally, every endpoint decision and generated rule must remain traceable to deterministic source evidence, supporting explainability ({\bf R4 - Inspectability and reproducibility}).

\subsection{Overview}   \label{subsec:Overview}
\name\ analyzes the source code of the \ac{iot} device software to produce the corresponding \ac{mud} file.
Rather than attempting exhaustive whole-program analysis, \name\ follows a \textit{best-effort, evidence-preserving} strategy that prioritizes repository regions likely to implement network behavior, and then progressively refines the recovered evidence.
As illustrated in Fig.~\ref{fig:AutoMUD-Overview}, the processing chain of \name\ includes four phases: 
\begin{figure*}[t!]
  \centering
  \includegraphics[width=\textwidth]{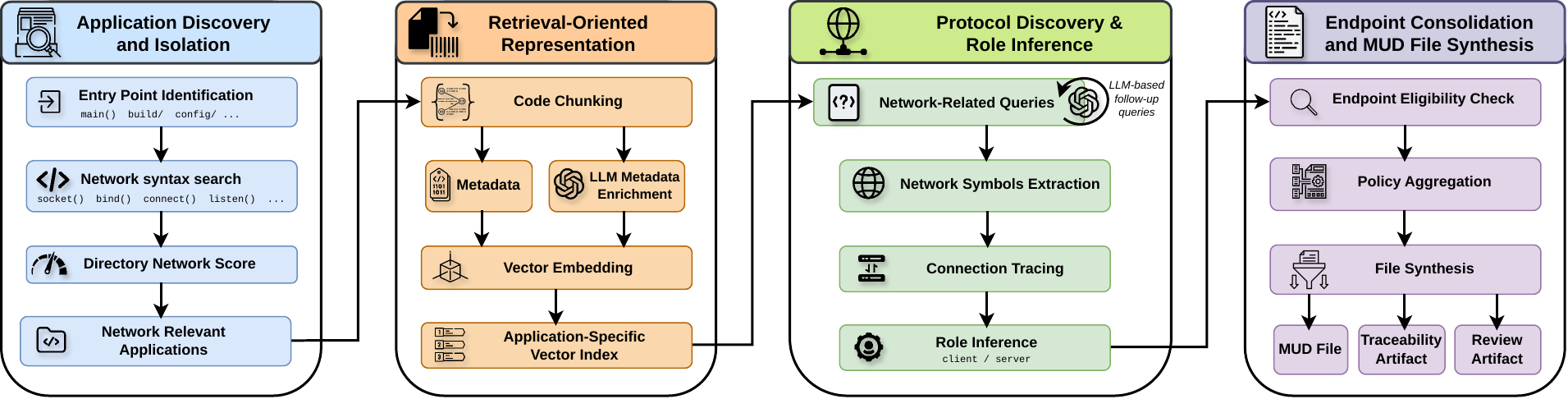}
  \caption{\name\ processing chain overview for automated \ac{mud} profile generation and traceability.}
  \label{fig:AutoMUD-Overview}
\end{figure*}
    First, \name\ performs \textbf{Application Discovery and Isolation}. Here, it identifies repository regions likely to implement application-level network behavior and groups them into independent candidate workspaces. This limits the analysis space and reduces contamination between unrelated components.
    Then, \name\ performs \textbf{Retrieval-Oriented Representation}. In this phase, source artifacts within each workspace are chunked, enriched with additive semantic metadata, embedded, and stored in an application-specific vector index.
    This processing enables subsequent stages to retrieve compact and traceable evidence.
    Next, \name\ conducts \textbf{Protocol Discovery and Role Inference}. It applies  semantic retrieval, syntactic extraction, and exact-source evidence to discover both constant-defined and behavior-defined communication capabilities.
    Through successive passes on the source code, the tool validates endpoint values and uses and infers the communication semantics required for policy generation . 
    Finally, \name\ conducts \textbf{Semantic Endpoint Consolidation and MUD File Synthesis}. It collects all the information generated by earlier processing steps, resolves inconsistencies or duplicates, and combines it into a structured knowledge base including standardized descriptions of the various endpoints.
    Through deterministic eligibility checks, \name\ selects records with concrete, source-supported policy semantics for compilation into the final \ac{mud} profile.
    Excluded or unresolved records remain available through traceability and operator-review artifacts.

Thus, together with the \ac{mud} profile corresponding to the analyzed codebase, \name\ also produces the traceability, validation, and review information required to inspect each policy decision. We describe the various phases in more details below.

\subsection{Application Discovery and Isolation}     \label{subsec:AppIsolation}
\ac{iot} software repositories commonly contain heterogeneous application logic, libraries, utilities, third-party components, configuration artifacts, and documentation~\cite{Dritsas:2025:IoTCyberSurvey, UlHaq:2023:FirmwareSurvey}.
Only part of this material contributes to the network behavior of a given device configuration.
Analyzing the complete repository uniformly would therefore be inefficient and could introduce unrelated evidence into downstream \ac{llm}-based reasoning.
This phase identifies repository regions likely to implement application-level network behavior and isolates them as independent analysis units.

Starting from the root of the traversable source tree, \name\ performs a lightweight, evidence-driven scan to construct candidate application workspaces.
It first identifies structural signals of executable or service-oriented code, including program entry points, executable scripts, build targets, and service initializers.
Non-traversable binary artifacts are excluded, whereas directory names and source categories influence discovery priority without determining inclusion.
This distinction prevents potentially executable network behavior from being discarded solely because it occurs in a test, compatibility, example, or other non-production subtree.
For each candidate region, \name\ estimates network relevance from lexical and syntactic indicators, including networking headers and APIs, protocol terminology, configuration declarations, and communication-oriented commands.
In C/C++ code, for example, relevant evidence includes networking headers and function calls such as \texttt{socket()}, \texttt{bind()}, \texttt{listen()}, and \texttt{connect()}~\cite{Stevens:2004:UNP,Kerrisk:2010:TLPI}.
These indicators are aggregated over repository subtrees to produce directory-level scores.
Rather than selecting every directory containing an isolated networking token, \name\ combines structural and scoring evidence to infer coherent application roots.
Each selected root becomes an independent workspace for subsequent indexing and analysis, reducing cross-component contamination and improving retrieval relevance~\cite{Xie:2023:DeepCodeSearchSurvey,li2024repofilter,Sourcegraph:2024:CodyContext}.

Discovery also attaches source-scope provenance to retained artifacts, distinguishing, when supported, production, test, regression, example, generated, documentation-related, and unresolved code.
Source scope is carried forward independently of behavioral semantics and does not by itself determine policy inclusion.
A non-production component that implements externally reachable communication under a supported configuration is evaluated under the same endpoint-eligibility criteria as production code.
Conversely, disabled, unreachable, loopback-only, or non-representable behavior remains ineligible.
Application isolation therefore determines where detailed analysis is performed, whereas endpoint eligibility later determines which recovered capabilities may become \ac{mud} rules.


\subsection{Retrieval-Oriented Representation}     \label{subsec:RetrievalRepresentation}
\name\ converts each candidate application into a fine-grained representation for semantic retrieval and \ac{llm}-based reasoning.
This phase produces an application-specific collection of enriched source chunks and a corresponding vector index, allowing later stages to retrieve compact, relevant evidence rather than process the complete workspace for every query.

\textbf{Code Chunking}.
For each application, \name\ builds an inventory of the readable artifacts and records deterministic file-level metadata, including the relative path, file type, language hint, size, and hash.
Files are then divided into bounded chunks suitable for subsequent reasoning~\cite{qu-etal-2025-semantic}.
Language-agnostic structural heuristics avoid evident improper boundaries, such as splitting a function, class declaration, or configuration block.
Each resulting record contains the raw source text, a stable identifier, canonical metadata, and source-line coordinates.
The latter allow later decisions to reference exact source regions, distinguish symbol definitions from their uses, and reconstruct evidence spanning adjacent chunks.

\textbf{Metadata Enrichment}.
Each chunk is enriched with metadata.
Metadata follows an explicit ownership contract.
Structural fields, including source path, file identity, chunk identifier, and line coordinates, are generated deterministically and remain authoritative throughout the pipeline.
The \ac{llm} receives bounded groups of chunks and may add network-oriented semantic annotations, such as protocol mentions, apparent roles, relevant functions, configuration fields, endpoints, dependencies, and network relevance.
These annotations must conform to a declared schema, remain grounded in the supplied text, and cannot overwrite structural metadata or establish unsupported numeric evidence.
The raw chunk text and its logical identity also remain unchanged if the \ac{llm}-facing context must be divided across multiple requests.
Before accepting the representation, \name\ verifies the canonical metadata contract and the correspondence between each record and its source material.
Incomplete or inconsistent representations are rejected.

\textbf{Vector Embedding and Application Indexing.}
\name\ then embeds the text of each chunk into a vector representation and stores it in an application-specific index.
Canonical and semantic metadata remain separate from the embedding and are retained for deterministic identity checks, filtering, ranking, and provenance.
At query time, the same embedding model represents the analysis query, and the nearest chunks are returned together with their metadata and source coordinates.
Maintaining a separate index for each candidate application prevents unrelated components from competing during retrieval and provides the source-traceable evidence consumed by protocol discovery and role inference.


\subsection{Protocol Discovery}  \label{subsec:ProtocolDiscovery}

\textbf{Adaptive Application Inspection.}
For each candidate application, \name\ first performs an exploratory retrieval pass to determine whether deeper protocol analysis is warranted.
This inspection allocates analysis effort but does not decide endpoint eligibility yet.
Broad queries, such as \textit{socket creation, connection setup, and protocol configuration}, retrieve an initial set of network-relevant chunks from the application index.
The results are deduplicated and ranked before the \ac{llm} produces a structured summary of the apparent modules, networking functionality, and available evidence.
For network-relevant applications, the \ac{llm} model generates follow-up queries targeting protocol use, socket operations, configuration, and communication roles.
These queries may reuse only identifiers observed in the retrieved evidence, allowing retrieval to adapt without introducing unsupported names.
For example, evidence containing HTTP configuration and socket handling may lead to queries for the observed connection functions, request handlers, and service-port symbols.
This refinement produces a compact context for candidate discovery and follows prior work showing that adaptive queries can improve retrieval quality~\cite{Lewis:2020:RAG,karpukhin-etal-2020-dense,Chan:2024:RQRAG}.


\textbf{Complementary Candidate Discovery.}
\name\ combines constant-oriented and behavior-oriented discovery because communication capabilities may be expressed either through named symbols or through configurable connection paths.
For constant-oriented discovery, network-relevant retrieval results and suitable source or header files undergo syntactic analysis.
A structured parser extracts functions and calls, field assignments, macros, constants, and enumeration values, reducing reliance on ad hoc regular expressions~\cite{Nielson:1999:PPA,Palix:2010:CoccinelleLinux,Yamaguchi:2014:CodePropertyGraphs}.
From this evidence, \name\ constructs an over-approximated inventory of protocol, address-family, socket-type, service, and port candidates required for later \ac{mud} synthesis~\cite{MUD-rfc,HernandezRamos:2021:MUDSurvey}.
Each candidate retains its symbol, source location, defining expression, and literal or symbolic value when available.
The \ac{llm} thus classifies these records and identifies repository-specific naming patterns, but the resulting inventory remains a set of source-grounded candidates rather than policy rules.

Behavior-oriented discovery targets capabilities represented through connection logic rather than fixed service constants, including configurable listeners, proxies, forwarding paths, and other build- or runtime-supplied endpoints.
Queries are constructed from application summaries, observed identifiers, configuration keys, and networking operations already present in the evidence.
Discovered candidates use the same record schema and later eligibility gates as constant-discovered candidates.
Configured, dynamic, and derived endpoints remain classified as such since network relevance alone cannot convert them into fixed ports.
Consequently, \name\ can retain these capabilities for explanation and review even when their endpoint semantics cannot be compiled automatically.



\textbf{Source-Authoritative Value Normalization.}
The high-recall inventory may include limits, sentinels, internal enumerations, configuration values, or unrelated numbers alongside genuine ports.
\name\ therefore separates semantic relevance from numeric authority.
The \ac{llm} may classify a symbol as networking-related, but its value becomes a port only when resolved deterministically from source evidence.
Fixed ports are accepted from integer literals or pure alias chains.
For example, if \texttt{PORT\_RTMPT} aliases \texttt{PORT\_HTTP\_TUNNEL}, which is defined as \texttt{80}, both symbols retain the source-resolved value 80.
General arithmetic and macro expressions are not evaluated into guessed ports.
Instead, values remain classified according to their evidenced form, such as configured, derived, range boundary, sentinel, internal enumeration, unresolved, or non-port.
Only source-verifiable fixed values or validated ranges may become packet-level port specifications; unsupported model-proposed values are ignored.

\textbf{Symbol-Use, Activation, and Port Validation.}
Because semantic retrieval may locate a candidate definition without recovering all of its uses, \name\ supplements retrieved evidence with deterministic exact-symbol search.
Word-boundary occurrences, neighboring chunks, and direct source context reveal whether a symbol is assigned, compared, passed to a networking operation, used in configuration logic, or propagated into an address structure.
This evidence supports subsequent interpretation while preventing relevant uses that cross chunk boundaries from being omitted.

\name\ also retains compile-time and runtime guards surrounding the recovered paths.
Preprocessor conditions and visible configuration checks distinguish behavior that is enabled, conditional, disabled, or unresolved under the available evidence.
A capability is conditional when a concrete network path exists but requires a build or runtime option; it is considered disabled only when the analyzed source or supplied configuration establishes that the path cannot execute.

Finally, a numeric candidate is accepted as a port only when its uses support that interpretation.
Positive evidence includes propagation into port-related fields, socket-address structures, or operations such as \texttt{htons()} and \texttt{getservbyport()}.
Conversely, use as a switch label, dispatch-table index, or type discriminator contradicts a fixed-port interpretation.
A candidate is classified as a non-port when independent evidence establishes such a control use, or as unresolved when the available evidence is insufficient; in both cases, the proposed value is retained for provenance.
This semantic classification remains separate from \ac{mud} eligibility, ensuring that even an already excluded candidate cannot preserve an unsupported fixed-port interpretation.


\subsection{Role Inference}

\textbf{Source-Grounded Connection Tracing}.
After discovering protocol and endpoint candidates, \name\ determines whether the device acts as a client, server, or both, providing the direction semantics required by the \ac{mud} model~\cite{MUD-rfc}.
This phase approximates control- and data-flow reasoning through iterative semantic retrieval, syntactic extraction, and \ac{llm}-based interpretation~\cite{Nielson:1999:PPA,Wang:2024:LLMDFA,Li:2024:IRIS,Salih:2025:LLMStaticSurvey}.
For each candidate, the analysis begins with its symbol, source-resolved value, exact occurrences, and neighboring code.
A parser extracts the relevant functions, structures, fields, guards, and source locations.
The \ac{llm} identifies the modules or handlers that own the behavior and generates bounded follow-up queries using only identifiers observed in the retrieved evidence.

Successive passes then trace endpoint values through socket-address fields, connection logic, and standard or project-specific networking operations~\cite{Stevens:2004:UNP,Kerrisk:2010:TLPI}.
Propagation to a remote-port field followed by \texttt{connect()}, for example, supports a client role, whereas assignment to a local address followed by \texttt{bind()} and \texttt{listen()} supports a server role.
Independently evidenced paths retain both roles, and local listening ports remain distinct from remote service ports.
Each decision records the supporting source regions, socket operations, propagation evidence, configuration conditions, and network semantics.
Role inference is therefore an incremental, source-grounded process rather than a single \ac{llm} judgment.

\textbf{Role and Context Separation.}
Each finding is normalized into independent semantic dimensions.
The \texttt{role} field records \texttt{client}, \texttt{server}, \texttt{both}, or \texttt{uncertain}, while \texttt{source\_scope} separately records the provenance of the code, such as production, test, regression, example, or unresolved.
Transport, IP family, endpoint side, configuration state, reachability, and supporting evidence are likewise represented independently.
Thus, a test server is encoded as \texttt{source\_scope=test} and \texttt{role=server}, rather than as a compound role.
This separation preserves provenance while allowing the later eligibility stage to determine whether the behavior is externally exercisable and representable in a \ac{mud} policy.

\textbf{Transport Evidence Consolidation.}
Transport-protocol evidence may be distributed across handler discovery, connection setup, socket assignment, address population, and port-use analysis.
Before endpoint classification, \name\ deterministically aggregates these observations into a per-candidate semantic ledger.
Strong TCP evidence includes markers such as \texttt{SOCK\_STREAM} and \texttt{IPPROTO\_TCP}.
Strong UDP evidence includes \texttt{SOCK\_DGRAM}, \texttt{IPPROTO\_}\\\texttt{UDP}, and datagram operations such as \texttt{sendto()} or \texttt{recvfrom()}.
Each observation retains a stable identifier, source phase and location, bounded excerpt, and indication of whether it occurs in direct network context.

Evidence without candidate-local network or port context is retained but given lower priority.
A transport protocol is considered corroborated only through direct strong evidence from independent phases or multiple independent implementation markers within one phase.
Even corroborated evidence does not automatically create an endpoint: the final decision must establish that the transport applies to the specific candidate rather than to unrelated infrastructure in the surrounding code.
The ledger therefore preserves transport evidence across phases without binding generic implementation markers to the final policy.



\subsection{Semantic Endpoint Consolidation and MUD File Synthesis} \label{subsec:MUDSynthesis}
The final phase consolidates the artifacts produced during protocol discovery, value normalization, and role inference into endpoint-level records.
For each candidate, the final classifier receives its source definition and resolved value, exact-use evidence, connection traces, configuration state, inferred roles, and transport ledger.
The classifier operates under a constrained schema and must ground every conclusion in the supplied evidence.
Its accepted outputs are normalized records that can be evaluated independently of the preceding \ac{llm} reasoning and, when eligible, compiled into \ac{mud} rules.

\textbf{Transport Applicability and Reconciliation.}
For every candidate, the classifier accounts explicitly for both TCP and UDP by assigning each transport one of four dispositions.
\texttt{represented} indicates that the transport applies to one or more returned endpoints and therefore requires endpoint indexes and supporting evidence.
\texttt{context\_only} retains genuine transport evidence that does not apply to the candidate, together with evidence identifiers and an explanation, but cannot reference an endpoint.
\texttt{uncertain} denotes insufficient evidence and triggers reconciliation, whereas \texttt{not\_observed} records the absence of transport evidence.
These schema constraints prevent previously collected evidence from being omitted or converted into an endpoint without justification.

When corroborated evidence is classified as \texttt{context\_only}, the explanation must relate its non-applicability to the candidate's symbol, source-resolved value, and surrounding source path.
An \texttt{uncertain} disposition triggers one dedicated reconciliation pass over the same source-grounded evidence.
This pass may resolve only the disputed transport applicability and cannot alter unrelated endpoint fields.
If uncertainty persists, the candidate remains available for traceability and review but is ineligible for automatic synthesis.
Transport ambiguity therefore cannot be resolved implicitly by the compiler.

\textbf{Endpoint Eligibility.}
Accepted classifications are normalized into endpoint records with stable identities and explicit evidence references.
These records are the only inputs consumed by the deterministic \ac{mud} compiler.
Eligibility requires three classes of conditions: packet-level semantics must be concrete and internally consistent, including a fixed port or source-validated range; the behavior must be enabled or conditionally supported, externally reachable, and representable by \ac{mud}; and its source values, port use, provenance, and final classification must pass deterministic validation and reconciliation.
These gates separate discovery from authorization.
Unresolved or non-representable capabilities remain visible in the analysis and review artifacts but are neither generalized into permissive rules nor emitted automatically.

\textbf{Policy Compilation.}
Eligible records from all applications are consolidated into a unified representation of device network behavior and merged only when their evidence establishes behavioral equivalence.
Thus, identical port numbers remain distinct when their transport, role, or activation conditions differ.
The compiler maps outbound client behavior to the \texttt{from-device-policy}, inbound listening behavior to the \texttt{to-device-policy}, and independently supported bidirectional behavior to both.
Endpoint-side semantics determine whether local and peer service ports become source- or destination-port matches, while stable record properties determine readable and traceable \ac{acl} and \ac{ace} names.
The resulting representation is serialized into the required \ac{mud} metadata, policies, \acp{acl}, and \acp{ace}; excluded records remain associated with their evidence and exclusion reasons for audit and review.


\subsection{Explainability, Traceability and Validation}    \label{subsec:explainability-validation}
\name\ produces three output classes.
The first is the standard-compliant \ac{mud} file containing the generated network policy.
The second comprises traceability and validation artifacts that explain each emitted rule and verify its consistency.
The third comprises review artifacts describing source-supported capabilities that were not eligible for automatic compilation.
Separating these outputs allows \name\ to remain conservative without concealing excluded or unresolved evidence.

\textbf{Source-to-Rule Traceability.}
Every generated \ac{ace} is linked to the normalized endpoint record from which it was compiled.
That record preserves the corresponding source origin, resolved values, behavioral evidence, semantic decisions, and eligibility outcome.
Intermediate artifacts capture the progression from protocol discovery and connection tracing to transport reconciliation, endpoint eligibility, and policy direction.
Consequently, explanations reconstruct each rule from the evidence used to generate it rather than from a post hoc interpretation of the final policy.

\textbf{Review of Excluded Capabilities.}
When a candidate fails one or more eligibility gates, the review output retains its normalized record, source provenance, coded exclusion reasons, and recommended verification actions.
If its packet-level semantics are complete, \name\ may also provide a non-authoritative candidate \ac{ace} for consideration after external verification.
No preview is produced when missing semantics would need to be inferred.
Review artifacts therefore expose potentially relevant capabilities without converting uncertainty into an enforceable permission.

\textbf{Structural and Cross-Artifact Validation.}
Before finalization, \name\ validates the structure and internal consistency of the generated \ac{mud} document.
The checks cover required metadata, policy and \ac{acl} references, \ac{ace} structure, match and action fields, address-family consistency, transport and port representation, and directionality.
The document is additionally validated against the RFC~8520~\cite{MUD-rfc} and RFC~8519 YANG modules~\cite{Watrobski:2022:NISTIR8349}.

Cross-artifact validation verifies traceability and completeness in both directions.
Every generated \ac{ace} must map to an eligible endpoint record and its supporting evidence, and every eligible record must map to the corresponding generated rules.
Conversely, each excluded record must appear in the review output with at least one explicit exclusion reason.
Review artifacts remain non-authoritative and cannot add, remove, or modify rules in the automatically generated policy.

\subsection{Prompt Engineering} \label{sec:PromptEngineering}
Because \ac{llm} outputs feed subsequent analysis stages, \name\ treats prompts as constrained interfaces between source evidence and structured artifacts.
Each prompt requires JSON-only output under a task-specific schema.
Reasoning is restricted to retrieved code, extracted identifiers, and previously accepted artifacts; adaptive queries may reuse only identifiers already present in the evidence; and classification tasks use closed label sets.
A model response becomes an analysis artifact only after satisfying the syntactic, schema, and semantic contract of its phase.

Prompts also preserve consistency across phases.
Source-validated port specifications remain authoritative, and every observed TCP or UDP marker must receive an explicit disposition, namely, \texttt{represented}, \texttt{context\_only}, \texttt{uncertain}, or \texttt{not\_observed}, with the required evidence references.
Corroborated evidence cannot be silently omitted, and distinct transport or activation conditions cannot be merged without support.
Unresolved applicability triggers a reconciliation pass restricted to the disputed decision; if uncertainty persists, the candidate is retained for review but excluded from \ac{mud} synthesis.

\section{Experimental Evaluation}   \label{sec:ExpEval}
We first present the experimental setup in Sec.~\ref{subsec:Setup}. Then, we evaluate \name\ through two complementary experimental campaigns.
The first assesses the effectiveness of the tool, i.e., whether it can recover source-grounded network behavior and translate it into a structurally valid and behaviorally consistent \ac{mud} profile (Sec.~\ref{subsec:EndToEnd}).
The second assesses the utility of the tool, i.e., 
it evaluates whether an erroneous source-to-policy decision that reaches the final profile can be detected, explained, and corrected using only 
the artifacts generated by \name\ (Sec.~\ref{subsec:fault-injection-evaluation}). We also compare the performance of \name\ to traffic-based approaches like MUDgee in Sec.~\ref{subsec:comparison}.

\subsection{Experimental Setup} \label{subsec:Setup}

\textbf{Codebase}.
In line with prior work on \ac{iot} security and static software analysis~\cite{Palix:2010:CoccinelleLinux,Ferrara:2021:StaticIoT,UlHaq:2023:FirmwareSurvey,Alrawi:2019:SoK,Dritsas:2025:IoTCyberSurvey,HernandezRamos:2021:MUDSurvey}, we construct a dedicated repository combining selected portions of a Raspberry Pi Linux kernel source tree with three widely used user-space networking programs: \texttt{curl}, \texttt{wget}, and \texttt{OpenSSH}. We select these applications because they encode network behavior through substantially different source-code organizations and declaration patterns, thereby exercising different capabilities of \name. For \texttt{curl} and \texttt{wget}, protocol definitions are concentrated in centralized header files and enumerations, enabling direct evaluation of symbol recovery. \texttt{OpenSSH} presents a more challenging setting, as endpoint interpretation requires combining evidence distributed across configuration processing, client and server connection setup, forwarding logic, channel management, and auxiliary utilities. Consequently, the selected codebase allows us to evaluate both networking-artifact recovery and semantic interpretation of communication behavior. 
Together, these applications provide the protocol, transport, port, role, and direction abstractions required for the generation of \ac{mud} profiles~\cite{Alrawi:2019:SoK,Mazhar:2021:MUDSurvey,HernandezRamos:2021:MUDSurvey}.

\textbf{Experimental Configuration.}
\emph{Embedding and Retrieval Configuration.} 
We use Haystack to orchestrate document storage and semantic retrieval, and we employ \texttt{intfloat/e5-small-v2} as the dense embedding model~\cite{Haystack:2025:Docs}\footnote{Model card: \url{https://huggingface.co/intfloat/e5-small-v2}.}. We only embed chunk text, while the canonical metadata, including repository identity, source path, chunk identifier, and source coordinates, are retained separately in an \texttt{InMemoryDocumentStore}.
Analysis queries are represented in the same embedding space and processed through Haystack's \texttt{InMemoryEmbeddingRetriever}, which returns the most relevant chunks together with their similarity scores and associated metadata. This separation preserves deterministic source identity and provenance throughout protocol discovery and role inference.

\emph{LLM Allocation.} 
We configure the \name\ pipeline to use three pinned OpenAI model snapshots, allocated according to task complexity and the potential impact of an incorrect decision~\cite{OpenAI:Models}.
We use \texttt{gpt-4o-mini} (\texttt{gpt-4o-mini-2024-07-18}) with \texttt{temperature = 0.0} for high-volume and structurally constrained tasks, including query expansion, constant extraction, value normalization, and schema-oriented JSON repair. This configuration promotes stable outputs, although it does not guarantee bit-for-bit reproducibility across independent executions.
We use \texttt{gpt-5-mini} (\texttt{gpt-5-mini-}\\\texttt{2025-08-07}) with \texttt{reasoning\_effort="low"} for repeated contextual analyses, including protocol exploration, connection tracing, and intermediate reconciliation tasks.
Finally, we use \texttt{gpt-5.1} (\texttt{gpt-5.1-2025-11-13}) with \texttt{reasoning\_effort="medium"} for \\repository-level interpretation, protocol identification, communication-role inference, endpoint classification, and final semantic reconciliation. Through this allocation, \name\ concentrates reasoning resources on the decisions most likely to influence the content of the generated \ac{mud} profile.

{\bf Ground Truth Construction.} We construct the evaluation ground truth manually and independently of the outputs produced by \name. To construct our ground truth, we examine selected source-code revisions together with their build and configuration artifacts and use official project documentation to corroborate externally visible protocol behavior~\cite{curl:Protocols,GNUWget:Manual,OpenSSH:sshdConfig}. Our choice of using a manually curated reference set is common in program-analysis evaluations when independent machine-readable specifications of complex software are unavailable~\cite{Nielson:1999:PPA,Tukaram:2022:SAWarnings,Ayewah:2007:FindBugsEval}.

We distinguish a retrieval-level ground truth, used to evaluate the recovery of networking artifacts, from a policy-level ground truth, used to evaluate endpoint interpretation and \ac{mud} synthesis.

\emph{Retrieval-Level Ground Truth.} The retrieval-level ground truth contains the networking artifacts implemented in the selected codebase that are relevant for subsequent endpoint analysis.
For \texttt{curl}, the reference comprises the 24 protocol identifiers defined in \texttt{urldata.h} and used during URL parsing and protocol dispatch. For \texttt{wget}, the reference comprises the three URL schemes supported by the selected configuration and declared in \texttt{url.h}. For \texttt{OpenSSH}, the reference contains four networking-related symbols relevant to externally visible communication behavior, including the standard SSH service port and the default ports associated with optional proxy functionality.
Overall, the retrieval-level ground truth contains 31 networking-related symbols: 24 from \texttt{curl}, 3 from \texttt{wget}, and 4 from \texttt{OpenSSH}. The objective of this stage is to evaluate whether \name\ can recover the networking artifacts implemented within the repository. It does not assess whether a recovered value corresponds to a deployable communication endpoint. Endpoint eligibility, communication semantics, and policy representation are evaluated separately through the policy-level ground truth.

\emph{Policy-Level Ground Truth.} Recovery of a networking artifact does not by itself establish whether the artifact represents a deployable communication endpoint or how it should appear in a MUD profile. Therefore, we inspect declarations, configuration paths, and call sites to determine each recovered behavior's transport protocol, port semantics, communication role, communication direction, activation conditions, and endpoint eligibility.

We consolidate artifacts referring to the same communication behavior into a single semantic representation, while range boundaries, internal discriminators, configuration markers, and other non-endpoint values are excluded from policy synthesis.

This process yields 23 source-validated semantic endpoint groups. Following the policy-construction methodology adopted in our evaluation, we represent each endpoint group for both IPv4 and IPv6 and for both initiating and complementary communication directions. Consequently, each endpoint group produces four address-family- and direction-specific \acp{ace}, yielding a reference policy containing \(23 \times 4 = 92\) \acp{ace}.

\subsection{End-to-End MUD Synthesis Evaluation}    \label{subsec:EndToEnd}
In this section, we evaluate to what extent \name\ recovers source-grounded network behavior and translates it into a structurally valid and behaviorally consistent \ac{mud} profile. We assess the tool at two complementary levels. 
Table~\ref{tab:end_to_end_results} summarizes our results.

\begin{table*}[t] 
\centering 
\caption{End-to-end MUD synthesis results on the evaluated repository.}
\label{tab:end_to_end_results} 
\small 
\begin{tabular}{l c c c c c c} 
\toprule \textbf{Application} & \textbf{Ground-Truth} & \textbf{Recovered} & \textbf{Ground-Truth} & \textbf{Recovered} & \textbf{Ground-Truth} & \textbf{Generated} \\ & \textbf{Symbols} & \textbf{Symbols} & \textbf{Endpoints} & \textbf{Endpoints} & \textbf{ACEs} & \textbf{ACEs} \\ \midrule curl & 24 & 24 & \multirow{3}{*}{23} & \multirow{3}{*}{23} & \multirow{3}{*}{92} & \multirow{3}{*}{92} \\ wget & 3 & 3 & & & & \\ OpenSSH & 4 & 4 & & & & \\ \midrule \textbf{Total} & \textbf{31} & \textbf{31/31} & \textbf{23} & \textbf{23/23} & \textbf{92} & \textbf{92/92} \\ \bottomrule 
\end{tabular} 
\end{table*}

\textbf{Network-Information Retrieval.}
We first evaluate whether \name\ can recover networking-related artifacts from the analyzed source code and distinguish deployable endpoints from networking-related but non-deployable values.
As summarized in Table~\ref{tab:end_to_end_results}, \name\ recovers all networking symbols contained in the reference repository: 24/24 for \texttt{curl}, 3/3 for \texttt{wget}, and 4/4 for \texttt{OpenSSH}. 
For \texttt{curl} and \texttt{wget}, all recovered symbols correspond to protocol identifiers declared in centralized header definitions. For \texttt{OpenSSH}, the recovered symbols include the standard SSH service port together with the SOCKS and HTTP proxy ports associated with optional gateway functionality. \texttt{IPPORT\_RESERVED} is also recovered, but excluded later because it is not a deployable endpoint (see discussion of the representative use case below). These findings demonstrate the ability of \name\ to identify communication behavior distributed across multiple repository components rather than concentrated in a single declaration location.

Recovering a networking-related symbol does not automatically imply that the corresponding value should appear in the generated policy. \name\ therefore performs additional semantic validation before endpoint synthesis. In particular, networking-related constants such as \texttt{IPPORT\_RESERVED} are correctly retained as contextual evidence while excluded from policy generation because they define implementation semantics rather than deployable communication endpoints.

Overall, the retrieval results show that \name\ can recover the networking artifacts required for policy generation while preserving the distinction between communication endpoints and non-endpoint implementation values.

\textbf{MUD Policy Synthesis.}
We next evaluate whether the recovered source-level behavior is translated correctly into a \ac{mud} profile.
The generated document satisfies the structural requirements of RFC 8520 and passes all validation checks described in Sec.~\ref{subsec:explainability-validation}. At the semantic level, \name\ correctly consolidates the recovered evidence into the 23 source-validated endpoint groups identified during ground-truth construction.
All 23 endpoint groups are represented in the generated policy, yielding 23/23 endpoint coverage. Following the reference-policy construction procedure, each endpoint group appears once for IPv4 and IPv6 and once for each traffic direction. Consequently, the generated profile contains \(92\) \acp{ace}, matching the reference policy exactly.
For \texttt{curl} and \texttt{wget}, \name\ correctly identifies the applications as communication initiators and generates the corresponding from-device and complementary return-traffic rules with appropriate transport and port semantics. For \texttt{OpenSSH}, the generated profile correctly preserves both TCP/22 client and server behavior and the conditional proxy-related communication capabilities identified during analysis. At the same time, networking-related but non-deployable values such as \texttt{IPPORT\_RESERVED} are excluded from policy synthesis and therefore do not appear as service permissions in the final profile.

Overall, the generated policy matches the manually derived reference policy at the semantic-rule level, preserving the expected transport protocol, port semantics, communication role, directionality, IP family, and activation context of every eligible endpoint.

\textbf{Traceability and Exclusion Correctness.}
We also evaluate the correctness of the traceability and review artifacts.
All 92 generated ACEs are linked to their originating endpoint records and supporting source evidence, yielding 92/92 rule-level traceability coverage. The corresponding artifacts preserve the provenance of each generated rule together with the intermediate decisions that led to its inclusion.

The review artifacts likewise preserve the reasons associated with excluded candidates. For example, OpenSSH channel-type constants such as\texttt{SSH\_CHANNEL\_PORT\_LISTENER} and \texttt{SSH\_CHANNEL\_}\\\texttt{RPORT\_LISTENER}, whose values are 2 and 11 respectively, are retained as source-supported evidence but classified as internal discriminators rather than deployable service ports. Consequently, no TCP/2 or TCP/11 permissions appear in the generated profile.

Overall, the results show that \name\ generates the expected policy and also maintains complete visibility over the source-to-rule mapping and the rationale underlying candidate exclusions.


\textbf{Execution Time and Cost.}
We measured five separate, successfully completed end-to-end runs on the curated curl, wget, and OpenSSH corpus.
The repository occupied 1.4 GB. In each run, the pipeline recorded 5,144 files and produced 23,839 source chunks.
Mean wall-clock execution time was 9 h 48 min (median 9 h 04 min; range 7 h 57 min to 14 h 11 min).
Protocol analysis accounted for 85.5\% of the aggregate execution time across runs, followed by semantic embedding at 14.1\%.
The pipeline issued an average of 2,651 model requests per run and processed 59.79 million tokens on average, comprising 56.91 million input and 2.88 million output tokens.
Provider usage metadata was available for every recorded request.


{\bf Implications.} The results demonstrate that source-code analysis can recover networking behavior and translate it into policy-level abstractions without requiring device deployment or traffic collection. Across the evaluated repository, \name\ recovered all networking artifacts contained in the ground truth, correctly consolidated them into semantic endpoint groups, and generated the expected MUD policy while preserving complete source-to-rule traceability.
Moreover, the results show that recovering a networking-related artifact is not sufficient for policy generation. \name\ correctly distinguishes deployable communication endpoints from networking-related implementation values and includes only eligible behaviors in the resulting policy. As illustrated by the OpenSSH analysis, networking artifacts that contribute to protocol handling, configuration management, or implementation logic can be retained for traceability and review without being translated into permissions.
These findings suggest that automated source-code analysis can support both policy generation and policy review. Beyond reconstructing communication behavior, \name\ preserves the evidence required to justify policy decisions and exposes the reasoning behind both inclusions and exclusions. This capability can assist manufacturers in validating generated profiles, reviewing auxiliary or conditional communication paths, and maintaining policies as software evolves.

\subsection{Controlled Fault Injection and Explainability Evaluation}   \label{subsec:fault-injection-evaluation}
In this section, we investigate whether the artifacts generated by \name\ support the detection, localization, explanation, and correction of semantic errors that propagate into an otherwise structurally valid MUD profile.

{\bf Experimental Setup.} We conduct a controlled fault-injection campaign using as the baseline the source-validated policies and \ac{mud} profile obtained in Sec.~\ref{subsec:EndToEnd}. Starting from the baseline, we inject semantic faults after protocol discovery and endpoint extraction, modifying only the high-level interpretation of selected candidate policies while preserving the underlying source code evidence, traceability records, and validation artifacts. Consequently, all injected policies remain structurally valid, and any inconsistency arises exclusively from an erroneous semantic decision of the tool. 

We consider five representative fault scenarios. Three faults derive from incorrect reinterpretation of implementation-specific constants as TCP ports, one is related to an incorrect derivation of a port expression as a fixed endpoint, and one is related to an erroneous conversion of an error-code discriminator into a service port. These cases were selected because they reflect realistic situations in which a recovered numeric value is network-related but does not correspond to a deployable communication endpoint.
We also use an independent \ac{llm}-based analyst as a blinded reviewer of the evidence that \name\ would make available to an operator. Specifically, the independent analyst uses the \textit{OpenAI Sol} model with \texttt{reasoning\_effort="max"}. The corresponding system prompt and audit instructions are reported in Appendix~\ref{app:AnalystPrompt}.
For each faulted policy, we provide the blinded analyst with the same artifacts that would be available to a human reviewer: the generated \ac{mud} file, validation report, traceability report, and operator-review artifacts. We withheld the clean policy, injected fault description, and all fault metadata. Thus, the analyst should determine whether generated rules are semantically supported by the available evidence and, if necessary, identify the required corrections.
The evaluation considers four criteria: (i) detection, i.e., identification of at least one inconsistent rule; (ii) localization, i.e., identification of the offending endpoint candidate; (iii) semantic reconstruction, i.e., recovery of the correct source-supported interpretation; and finally, (iv) policy recovery, restoration of the original clean policy without affecting unrelated rules.

{\bf Results.} Table~\ref{tab:fault_campaign} summarizes the results of the fault-injection campaign.
\begin{table*}[t] 
\centering \caption{Results of the controlled fault-injection campaign.} \label{tab:fault_campaign} \small \begin{tabular}{l l c c c c c} \toprule \textbf{ID} & \textbf{Injected Fault} & \textbf{Affected} & \textbf{Detected} & \textbf{Localized} & \textbf{Reconstructed} & \textbf{Recovered} \\ & & \textbf{ACEs} & \textbf{ACEs} & \textbf{ACEs} & \textbf{ACEs} & \textbf{ACEs} \\ \midrule E01 & \texttt{SSH\_CHANNEL\_PORT\_LISTENER} interpreted as TCP port 2 & 4 & 4 & 4 & 4 & 4 \\ E02 & \texttt{SSH\_CHANNEL\_RPORT\_LISTENER} interpreted as TCP port 11 & 4 & 4 & 4 & 4 & 4 \\ E03 & \texttt{X11\_BASE\_PORT} interpreted as fixed TCP port 6000 & 8 & 8 & 8 & 8 & 8 \\ E04 & \texttt{PE\_BAD\_PORT\_NUMBER} interpreted as TCP port 4 & 4 & 4 & 4 & 4 & 4 \\ E05 & Additional semantic endpoint misclassification & 4 & 4 & 4 & 4 & 4 \\ \midrule \textbf{Total} & -- & 24 & \textbf{24/24} & \textbf{24/24} & \textbf{24/24} & \textbf{24/24} \\ \bottomrule \end{tabular}
\end{table*}
Using the artifacts generated by \name, the analyst correctly detected and localized all five injected faults and corresponding erroneous \acp{ace}. In every case, the analyst reconstructed the source-supported interpretation of the affected candidate and identified the appropriate modification to the generated policy. By applying the recommended corrections, the analyst restored the original source-validated policy without introducing collateral changes to unaffected rules.
Across the experiments, the injected faults produced between four and eight additional \acp{ace} per policy, depending on the specific endpoint interpretation error. After correction, all faulted policies returned to the clean 92-\ac{ace} baseline. No unaffected \acp{ace} were removed or modified.
These results indicate that the traceability and review artifacts generated by \name\ preserve sufficient information to reconstruct the reasoning behind policy-generation decisions and identify inconsistencies that are not detectable through structural validation alone.

{\bf Representative Example.} Among the injected cases, we consider the reference use case of the interpretation of the constant \texttt{SSH\_CHANNEL\_PORT\_LISTENER}.

In the clean analysis, \name\ correctly extracts the value \texttt{SSH\_CHANNEL\_PORT\_LISTENER = 2} but classifies it as an internal OpenSSH channel-type discriminator rather than a TCP port. Therefore, the corresponding candidate is excluded from \ac{mud} synthesis. To generate the fault, we modified the endpoint record to represent the value as a fixed TCP listener on port 2 and marked it as eligible for policy generation.

The resulting \ac{mud} profile remained structurally valid, but contained four additional TCP/2 \acp{ace}, one for each IPv4/IPv6 and policy-direction combination. However, the traceability artifacts still recorded the candidate as an internal discriminator, reported no positive evidence of port use, and linked the symbol exclusively to switch statements, dispatch tables, and channel-type comparisons.

Using only the generated artifacts, the analyst was able to trace the four affected \acp{ace} to their originating candidate, identified the inconsistency between the generated rules and the recorded evidence, and concluded that the value 2 represented an implementation-specific discriminator rather than a deployable service port. Therefore, the corresponding endpoint was reclassified as configuration-dependent and ineligible for inclusion into the \ac{mud} file. Recompiling the policy removed the four injected \acp{ace} and restored the original 92-rule profile without modifying any other entry.

The remaining fault cases followed a similar pattern. In each case, the analyst exploited the preserved provenance and validation evidence provided by \name\ to separate source-level extraction correctness from the higher-level semantic interpretation that ultimately generated the erroneous rule. More details are reported in Appendix~\ref{app:additional-fault-injections}.

{\bf Implications.} The results demonstrate a distinction between structural validity and semantic correctness. A \ac{mud} profile may satisfy all syntactic and schema-level requirements while still containing permissions that are unsupported by the underlying implementation evidence. Conventional validation mechanisms are generally unable to detect such inconsistencies because the generated \ac{mud} file remains formally correct.
By preserving the complete source-to-rule chain, \name\ enables post-generation auditing of policy decisions. Rather than treating traceability solely as an explainability feature, the fault-injection campaign shows that it can support practical error diagnosis and policy recovery when semantic faults occur. These findings suggest that source-grounded provenance can improve the accountability and maintainability of automatically generated \ac{mud} profiles, particularly when policies are reviewed by security analysts or manufacturers before deployment.

\subsection{Comparison to Traffic-based Approaches}
\label{subsec:comparison}


In this section, we compare \name\ to the closest approach to \ac{mud} profile generation in the current state of the art, i.e., network-based approaches such as MUDgee~\cite{hamza2022_tdsc,Hamza:2018:ClearAsMUD}.
Traffic-based tools such as MUDGee generate \ac{mud} profiles by observing network communications and translating the observed flows into policy rules. Such approaches can reduce the effort required for manual policy construction, but their visibility is inherently limited to communications that occur during the observation period. By contrast, \name\ derives network behavior directly from source code and is therefore not constrained by whether a communication pattern is exercised during monitoring.

To compare these two approaches, we evaluate the impact of various time-limited observation windows on \ac{mud} profile completeness. We consider a traffic-based \ac{mud} profile generator that can perfectly translate every observed communication into the corresponding \ac{mud} rule, and we investigate how likely it is for our solution and such a traffic-based approach to recover the complete \ac{mud} profile represented by the source code.
Note that this assumption intentionally favors traffic-based approaches: any omitted policy is attributable exclusively to the absence of traffic during observation rather than to errors in traffic processing, protocol identification, or policy generation. Thus, our analysis provides a worst-case comparison of \name\ against the entire class of traffic-based generators.

{\bf Experimental Setup.} For our analysis, we use the same source-validated policy obtained in Sec.~\ref{subsec:Setup}. The generated profile contains 92 \acp{ace} corresponding to 23 semantic endpoint groups. 
Thus, we use the $n=23$ semantic groups as the fundamental observation units.
We model the appearance of each communication group within an observation window $T$ as an independent Poisson process. A communication group is considered successfully recovered by the traffic-based \ac{mud} generator if at least one corresponding network event occurs during the observation period. 

We evaluate several normalized communication-rate scenarios. Let $\lambda_c$ denote the rate associated with commonly occurring communication groups and $x=\lambda_c \cdot T$ the normalized group exposure. To capture the presence of rare or infrequently exercised behaviors, we designate $r \in \{ 1, 2, 5 \}$ communication groups as low-rate groups. We define their activation rate as $\lambda_r= \alpha \cdot \lambda_c$, where $\alpha \in \{1,0.1,0.01,0.001\}$ defines how rare it is to observe a communication pattern in $T$. Therefore, the probability of observing the complete source-validated profile is: $
P_{\mathrm{complete}}(x)=
\left(1-e^{-x}\right)^{n-r}
\left(1-e^{-\alpha x}\right)^r.
$
For each configuration, we evaluate both the expected fraction of recovered communication groups and the probability of recovering the complete 23-group profile.

\begin{figure}
    \centering
    \includegraphics[width=\linewidth]{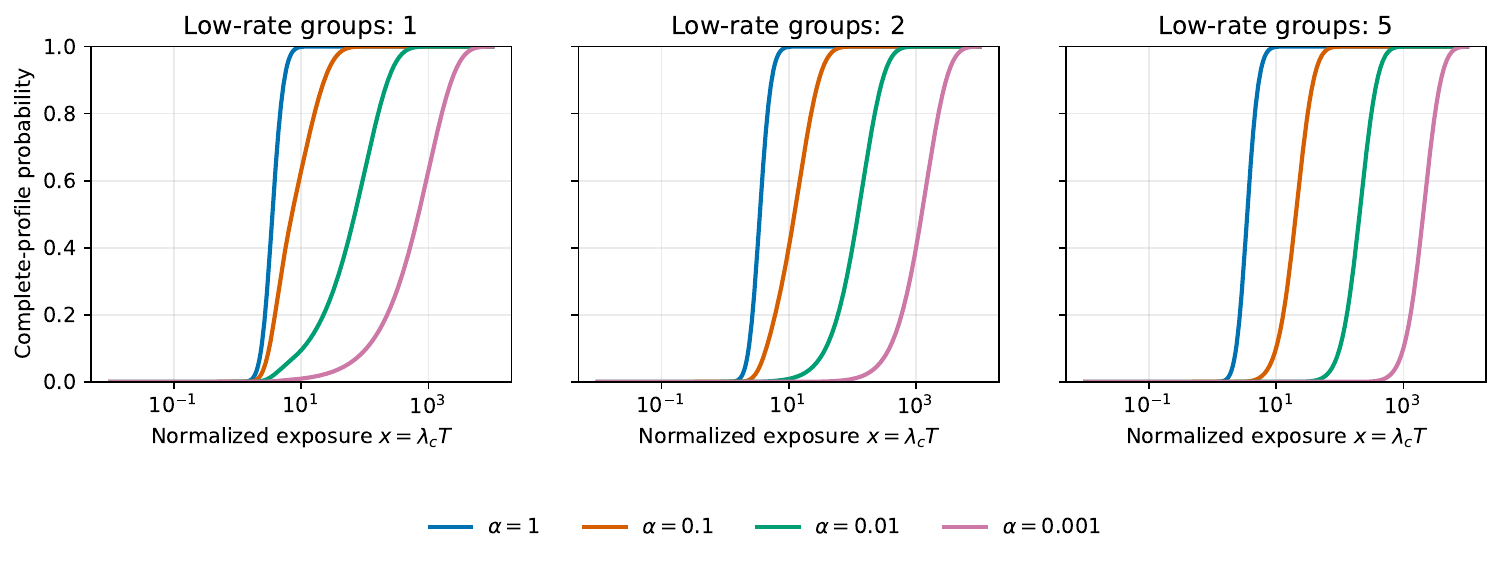}
    \caption{Probability of producing a complete \ac{mud} profile as a function of the normalized observation exposure with 1 (a), 2 (b) and 5 (c) low-rate groups.}
    \label{fig:comparison_compProb}
\end{figure}

{\bf Results.} Figure~\ref{fig:comparison_compProb} reports the probability of producing a complete MUD profile as a function of the normalized observation exposure with 1 (a), 2 (b) and 5 (c) low-rate groups. 

Figure~\ref{fig:comparison_compProb} shows that when all communication groups exhibit comparable activation frequencies ($\alpha=1$), the probability that network-based approaches recover the complete \ac{mud} profile increases rapidly with the observation exposure and eventually approaches $1$. However, introducing only a small number of infrequently exercised behaviors significantly changes the outcome. When a single communication group is ten times less likely to appear than the others ($\alpha=0.1$), substantially larger observation windows are required before the complete behavior defined in the source code becomes visible in network traffic. For $\alpha=0.01$ and $\alpha=0.001$, \ac{mud} profile completeness becomes dominated almost entirely by the rare communication group. This effect becomes even more pronounced when multiple low-rate groups are present.
To provide a reference example, considering an observation probability of $0.95$ for each common communication group, the probability of recovering the complete 23-group \ac{mud} profile is only $0.307$ when all groups occur at similar rates. Introducing a single low-rate group reduces this value to $0.838$, $0.0955$, and $0.00968$ for values of $\alpha$ of $0.1$, $0.01$, and $0.001$, respectively. 
Interestingly, the expected communication group coverage in all scenarios remains high, typically above $0.90$. Therefore, a traffic-based \ac{mud} profile generator may appear to have captured most of the device behavior while still missing one or more legitimate communications required to reconstruct the complete source-validated \ac{mud} profile.

{\bf Implications.} Overall, these results highlight the intrinsic limitations of traffic-based \ac{mud} generation. Even under idealized assumptions where every observed communication is translated perfectly into a policy rule, time-limited observation windows cannot guarantee complete visibility of implemented behavior, and the recovery process is ultimately governed by the least frequently exercised communication pattern.
By contrast, \name\ does not depend on runtime occurrence of communications. Once the relevant source code is available, the tool can recover communication capabilities regardless of whether they are exercised during a particular monitoring campaign. As a result, \name\ can expose configuration-dependent, rarely activated, or difficult-to-trigger behaviors that may remain absent from finite traffic traces.

More in general, our comparison suggests that source-code analysis and traffic observation provide complementary perspectives. Traffic traces reveal which communications are exercised in a particular deployment, whereas source-code analysis exposes the set of communication capabilities implemented by the software. Combining both sources of information represents a promising direction for future \ac{mud} generation workflows.

\section{Discussion}    \label{sec:Discussion}

\subsection{Key Findings}
Our experimental results support three main findings.

First, \emph{implementation evidence can be translated into traceable policy-level behavior}.
\name\ recovers all 31 symbols in the ground truth and translates the 24 source-validated semantic endpoint groups into the expected 92~\acp{ace}.
The generated profiles pass the implemented structural and cross-artifact checks, contain no duplicate rules, and link every emitted \ac{ace} to its endpoint record and supporting source evidence.
Thus, \name\ effectively bridges the semantic gap between low-level implementation facts and the protocol, transport, port, role, direction, and address-family information required by \ac{mud}.
As demonstrated by the comparison in Sec.~\ref{subsec:comparison}, this source-based perspective differs from traffic-based generators such as MudGee, whose rules can represent only communications exercised and captured during observation~\cite{Hamza:2018:ClearAsMUD,Hamza:2018:MUDgeeTool}.
Once the relevant source tree and configuration are available, \name\ can construct the policy without first deploying the device or waiting for each behavior to occur.
Consequently, it can recover rare, failure-triggered, update-related, or configuration-dependent paths that may remain absent even from long traffic traces.
For manufacturers, \name\ supports earlier profile construction from the versioned artifacts used to develop the device and provides a shared source-to-rule representation for firmware developers, product-security teams, and network-policy engineers.

Second, we observe that \emph{sources-driven analysis can expose overlooked capabilities without converting every network-related value into a permission}.
Our results specifically for OpenSSH illustrate both properties.
During initial investigation, \name\ helped us identifiying conditional SOCKS and HTTP proxy behavior not recognized at after a first iteration of our analysis, and manual investigation confirmed that the corresponding functions represent genuine, externally visible capabilities under the relevant configuration.
At the same time, \name\ distinguishes these endpoints from \texttt{IPPORT\_RESERVED} and the channel-type values 2 and 11, which are network-related implementation facts but not deployable services.
Thus, 
automated source analysis can systematically direct attention of a human expert to secondary or dispersed functionality that may be missed during manual reconstruction.
This is particularly important for code located in regression, test, compatibility, or optional execution paths.
In general, the existence of such functionalities does not itself establish that they should be authorized in a production policy.
In fact, they may be intentionally supported, enabled only under a specific configuration, accidentally included in a build, or left over from testing.
By preserving source scope, activation conditions, eligibility outcomes, and exclusion evidence, \name\ allows manufacturers to decide whether a networking capability should be retained, disabled, or represented explicitly, without silently turning every recovered value into an enforceable rule.
Overall, source analysis and traffic observation should be considered complementary approaches, rather than interchangeable.
The source code exposes all the supported networking capabilities, including dormant or conditional paths, whereas network traffic shows which capabilities are exercised by a particular deployment and workload.
Runtime observation could refine further a source-derived profile by confirming active flows, providing deployment-specific peer information, and distinguishing routinely exercised behavior from behavior that remains inactive in the monitored environment.
Future work in this direction could design and evaluate a combined workflow using source analysis to reduce the risk of omitting legitimate but infrequent communication and traffic observation to specialize the resulting policy for its deployment context.

Finally, \emph{traceability is particularly useful when a generated rule is wrong}.
As shown in Sec.~\ref{subsec:explainability-validation} across various representative fault cases, the injected semantic inconsistencies reached structurally valid policies, but remained inconsistent with the lower-level evidence retained by \name.
These results highlight a distinction between structural validity and semantic accountability.
A syntactically valid \ac{mud} file may still contain a plausible but unsupported permission.
By separating discovery, semantic classification, eligibility assessment, and deterministic compilation while preserving the related source-code evidence, \name\ prevents the final file from erasing contradictory provenance.
The retained artifacts can therefore support investigation of both emitted rules and capabilities withheld from automatic compilation.
Thus, thanks to \name, manufacturers and operators can focus the review on ambiguous or consequential decisions, without manual reconstruction of the complete source-to-policy path.

\subsection{Limitations and Future Work}    \label{subsec:discussion-limitations}
\name\ follows a best-effort analysis strategy and assumes access to a traversable source tree and relevant configuration evidence.
Missing proprietary components, generated code, external libraries, build-time values, or unretrieved source regions may conceal communication behavior. Future work could consider extending \name\ further to support additional information about unaccessible source regions, e.g., publicly-available information and traffic-based sources.

For evaluation purposes, we used a private-owned repository as an intentionally controlled benchmark combining mature networking applications with kernel-level material. This strategy is necessary for assessing quantitatively the performance of \name\ in a controlled setup. Future work could consider expanding the evaluation to complete device software trees, larger cross-vendor repositories, additional languages and build systems, and embedded or non-Linux environments. 

For evaluating the potential of \name\ in solving inconsistencies, we rely on a bounded set of higher-level semantic faults for which contradictory lower-level evidence is available. Although this is a limited set of the possible inconsistencies that may arise in this domain, it serves as a representative example of the potential of \name. Finally, the same evaluation relies on a blinded OpenAI Sol analyst. Future work could design a systematic user study involving human SoC operators.

\section{Conclusions}   \label{sec:Conclusions}
This paper presented \name, a source-code-driven approach for automatically generating \ac{mud} profiles from source code.
\name\ combines lightweight code analysis, retrieval-grounded \ac{llm} reasoning, and deterministic validation to translate implementation artifacts into policy-eligible network behavior without requiring device deployment or network traffic traces. Moreover, \name\ generates traceable policies: every emitted rule remains linked to the source evidence that produced it, while unresolved or non-representable findings are retained for further review.
Our evaluation demonstrates that \name\ recovers all source-grounded communication behaviors contained in the benchmark, generated the expected MUD policy, and preserved complete source-to-rule traceability.
Our controlled fault-injection campaign further showed that the produced artifacts can support the detection, explanation, and correction of semantic policy errors.

Overall, our work demonstrates that source-code analysis can complement traditional traffic-based \ac{mud} generation by exposing implemented capabilities that may remain unobserved during time-limited monitoring campaigns. This contributes a step forward toward reducing the effort of \ac{mud} adoption in the real-world and improving the transparency and maintainability of \ac{iot} network security policies.


\newpage



\bibliographystyle{ACM-Reference-Format}
\bibliography{Bibliography}

\appendix
\section{LLM Analyst Prompt}    \label{app:AnalystPrompt}
Listing~\ref{lst:fault-injection-prompt} shows the prompt issues to the LLM-based alayst for the controlled fault injection evaluation, while Listing~\ref{lst:fault-injection-example} shows the corresponding output.

\begin{lstlisting}[style=promptstyle, label=lst:fault-injection-prompt, caption=LLM-based analysist foult injection evaluation provided prompt.]
I provide you with a MUD file generated automatically from source-code analysis, together with the validation, traceability, and review artifacts produced by the generation pipeline.
Treat the generated MUD as potentially correct or potentially affected by an upstream analysis error; do not assume either outcome in advance.
Audit the final MUD against the supporting artifacts and determine whether all generated rules are internally consistent with their recorded evidence.
If you identify an inconsistency:
1. identify precisely the affected MUD ACE or ACEs;
2. trace them to the originating endpoint/candidate using the supplied artifacts;
3. explain which recorded evidence supports or contradicts the generated rule;
4. determine the most likely correct semantic interpretation;
5. reconstruct the endpoint representation that should have been used;
6. state precisely what change should occur in the final MUD;
7. identify whether any other MUD rules should change.

Base the analysis only on the supplied artifacts. Clearly distinguish what is directly evidenced from what is inferred. If the supplied artifacts are insufficient to complete any step, state exactly which additional artifact or evidence would be required.
\end{lstlisting}

\begin{lstlisting}[style=promptstyle, label=lst:fault-injection-example, caption=LLM-based analysist foult injection evaluation output result.] 
The generated MUD is structurally valid but not semantically consistent with its recorded evidence. Exactly four ACEs erroneously authorize TCP port 2 from `SSH_CHANNEL_PORT_LISTENER`, although the traceability artifact classifies `2` as an internal OpenSSH channel-type discriminator. The remaining 92 ACEs have no direct contradiction in the supplied artifacts.

1. Affected ACEs
All four originate from `SSH_CHANNEL_PORT_LISTENER-6e3952a5512a:endpoint:0`:
* `mud-to-device-ipv4`: `local-tcp-port-forward-listener-ch-to-d-tcp-5180e81533`
* `mud-from-device-ipv4`: `local-tcp-port-forward-listener-ch-from-tcp-24163b4716`
* `mud-to-device-ipv6`: `local-tcp-port-forward-listener-ch-to-d-tcp-212105e9bf`
* `mud-from-device-ipv6`: `local-tcp-port-forward-listener-ch-from-tcp-4152ff5387`

These ACEs accept IPv4/IPv6 TCP traffic using source or destination port 2 according to direction.
[...]

2. Origin and evidence
The traceability artifact maps all four ACEs to candidate `SSH_CHANNEL_PORT_LISTENER-6e3952a5512a` in `openssh-portable-master`, extracted from:
#define SSH_CHANNEL_PORT_LISTENER 2 /* Listening on a port. */

The recorded `raw_value` and `normalized_value` establish only that the macro value is 2; they do not establish that 2 is a network port.
The contradicting evidence is explicit:
* `network_port_usage_validation.status` is `internal_discriminator`;
* `positive_port_evidence` is empty;
* the symbol occurs as a case label, dispatch-array index, and channel-type comparison.

For example:
case SSH_CHANNEL_PORT_LISTENER:
pre[SSH_CHANNEL_PORT_LISTENER] = &channel_pre_listener;
post[SSH_CHANNEL_PORT_LISTENER] = &channel_post_port_listener;
is_client = (type == SSH_CHANNEL_PORT_LISTENER);

Thus, `2` identifies an internal channel type/state and is not passed to `bind()`, `listen()`, or another network-port operation. The upstream error was interpreting this macro value as a TCP port.
The separate review record `ssh_local_port_forward_L-e6fb24f53bd3:endpoint:0` confirms that the actual listener port is supplied through `-L` or `LocalForward` and represents it as configuration-dependent:
{
"kind": "configured",
"value": null,
"expression": null
}
[...]

3. Correct interpretation and reconstruction
Directly evidenced, `SSH_CHANNEL_PORT_LISTENER` denotes an OpenSSH channel type associated with a local forwarding listener, while its numeric value is an internal discriminator. The actual listening port is configuration-dependent.
The endpoint should therefore be rejected as a fixed-port candidate. If retained for auditability, its essential representation should be:
{
"endpoint_id": "SSH_CHANNEL_PORT_LISTENER-6e3952a5512a:endpoint:0",
"behavior": "SSH local TCP port-forward listener using a configured listening port",
"port": null,
"role": "server",
"endpoint_side": "local_listener",
"socket_action": "listen",
"transport": "tcp",
"ip_family": "both",
"port_spec": {
"kind": "configured",
"value": null
},
"configuration_state": "conditional",
"mud_eligible": false
}

Applying these configured-port semantics to this exact endpoint is an inference, but the same semantics are directly recorded for the corresponding `ssh_local_port_forward_L` endpoint.
[...]

4. Required correction
Remove the four TCP/2 ACEs. They must not be replaced with another fixed-port rule based on the supplied evidence.
After correction:
* total ACEs: 96 -> 92;
* ACEs per ACL: 24 -> 23;
* TCP/2 ACEs: 4 -> 0.

The policy and ACL names remain unchanged. A new rule could be generated only if deployment evidence later establishes a concrete forwarding port.

5. Other rules
No other generated ACE needs to change. The 96 MUD ACEs and 96 traceability records have exact name correspondence, and the other 92 ACEs agree with their recorded family, transport, direction, role, and numeric port evidence. No other included candidate is classified as an `internal_discriminator`.
The analogous `SSH_CHANNEL_RPORT_LISTENER = 11` candidate was correctly excluded, demonstrating how the port-2 candidate should have been handled. The 33 currently excluded endpoints should remain excluded; in particular, no TCP/11 rule should be introduced.
\end{lstlisting}

\section{Additional Controlled Fault-Injection Cases}    \label{app:additional-fault-injections}
Listing~\ref{lst:fault-injection-example} reports the complete output for E01, in which the value of \texttt{SSH\_CHANNEL\_PORT\_LISTENER} was incorrectly interpreted as TCP port 2.
The four remaining cases were evaluated using the same artifacts and analysis procedure.
Their complete reports followed the structure shown in the listing but are omitted to avoid repetitive material.
We instead summarize each injected error and the corresponding diagnosis below.

\textbf{Injection case - \texttt{SSH\_CHANNEL\_RPORT\_LISTENER}.}
This injection treated the value \texttt{SSH\_CHANNEL\_RPORT\_LISTENER = 11} as a fixed TCP port and introduced four TCP/11 ACEs.
The analyzer established that the symbol identifies an internal OpenSSH channel type used for remote-forwarding listeners.
Its value occurs in switch cases, channel-dispatch tables, and channel-type comparisons; it is not supplied to a socket operation as a port number.
The actual remote-forwarding port is obtained from runtime configuration, such as an \texttt{-R} or \texttt{RemoteForward} directive.
The analyzer therefore removed the four TCP/11 ACEs, reducing the faulted policy from 96 to the expected 92 ACEs.


\textbf{Injection case - \texttt{X11\_BASE\_PORT}.}
This injection converted \texttt{X11\_}\\\texttt{BASE\_PORT = 6000} into eight fixed TCP/6000 ACEs.
Unlike the preceding constants, this value participates in computing a network port; however, it is only the base of the X11 port calculation.
OpenSSH derives the effective listener port as \texttt{X11\_BASE\_}\\\texttt{PORT + display\_number}.
The source artifacts did not establish a fixed display number or prove that the resulting deployment port was always 6000.
The analyzer therefore rejected the fixed-port interpretation and removed the eight unsupported ACEs, reducing the policy from 100 to 92 ACEs.
A TCP/6000 rule would be justified only if separate configuration evidence established display number zero for the analyzed deployment.

\textbf{Injection case - \texttt{PE\_BAD\_PORT\_NUMBER}.}
The final injection interpreted \\\texttt{PE\_BAD\_PORT\_NUMBER = 4} as TCP port 4 and generated four corresponding ACEs.
The analyzer identified the symbol as a URL-parser error-code discriminator indicating that a supplied port is malformed or outside the valid range.
The value selects an error condition; it does not identify the remote port contacted by the application.
The actual destination port is instead obtained from the parsed URL or the relevant protocol default.
Accordingly, the analyzer removed the four TCP/4 ACEs and restored the policy from 96 to 92 ACEs.

Across all these four cases, the analyzer identified the injected symbol, traced the affected ACEs to their originating candidates, explained why the extracted numeric value did not justify the generated rule, and prescribed the expected correction
Each corrected policy matched the clean 92-ACE baseline, and no unaffected ACE was removed or modified.

\end{document}